\documentclass[aps,amsmath,twocolumn,amssymb,floatfix,showpacs,superscriptaddress,longbibliography]{revtex4-1}
\usepackage{amsmath}
\usepackage{mathtools}
\usepackage{braket}
\usepackage{bm}
\usepackage{graphicx}
\usepackage{amssymb}
\usepackage[dvipsnames]{xcolor}
\usepackage{multirow}
\usepackage{xfrac}
\usepackage{textcomp}
\usepackage{float}
\usepackage{enumerate}
\usepackage{subfigure}

\usepackage[colorlinks=true,linktoc=page,linkcolor=ForestGreen,citecolor=blue,urlcolor=purple]{hyperref}
\usepackage[toc,page]{appendix}

\newcommand{\ie}{{\it i.e.,\,\,}}

\newcommand\bea{\begin{eqnarray}}
\newcommand\eea{\end{eqnarray}}
\newcommand\beq{\begin{equation}}  
\newcommand\eeq{\end{equation}}

\usepackage[normalem]{ulem}
\definecolor{lime}{HTML}{A6CE39}
\usepackage{sidecap,tikz}
\DeclareRobustCommand{\orcidicon}{\hspace{-1.0mm}
	\begin{tikzpicture}
	\draw[lime, fill=lime] (0.0,0.0) 
	circle [radius=0.15] 
	node[white] {{\fontfamily{qag}\selectfont \tiny \,ID}};
	\draw[white, fill=white] (-0.0525,0.095) 
	circle [radius=0.007];
	\end{tikzpicture}
	\hspace{-3.0mm}
}
\foreach \x in {A, ..., Z}{\expandafter\xdef\csname orcid\x\endcsname{\noexpand\href{https://orcid.org/\csname orcidauthor\x\endcsname}
		{\noexpand\orcidicon}}
}

\begin{document} 


\title{Non-local spin entanglement in a fermionic chain }

\author{Sayan Jana \orcidB{}}
\email{sayan@iopb.res.in}
\affiliation{Institute of Physics, Sachivalaya Marg, Bhubaneswar-751005, India}
\affiliation{Homi Bhabha National Institute, Training School Complex, Anushakti Nagar, Mumbai 400094, India}

\author{Anant V. Varma \orcidD{}}
\email{anantvijay.cct@gmail.com}
\affiliation{Indian Institute of Science Education and Research Kolkata, Mohanpur, Nadia 741246, West Bengal, India}

\author{Arijit Saha\orcidC{}}
\email{arijit@iopb.res.in}
\affiliation{Institute of Physics, Sachivalaya Marg, Bhubaneswar-751005, India}
\affiliation{Homi Bhabha National Institute, Training School Complex, Anushakti Nagar, Mumbai 400094, India}

\author{Sourin Das\orcidA{}}
\email{sourin@iiserkol.ac.in; sdas.du@gmail.com}
\affiliation{Indian Institute of Science Education and Research Kolkata, Mohanpur, Nadia 741246, West Bengal, India}

\begin{abstract}

 An effective  two-spin density matrix (TSDM) for a pair of spin-$1/2$ degree of freedom, residing at a distance of $R$ in a spinful Fermi sea, can be obtained from the two-electron density matrix following the framework prescribed in Phys. Rev. A {\bf 69}, 054305 (2004). We note that the single spin density matrix (SSDM) obtained from this TSDM for generic spin-degenerate systems of free fermions is always pinned to the maximally mixed state $i.e.$ $(1/2)  \ \mathbb{I}$, independent of the distance  $R$ while the TSDM confirms to the form for the set of maximally entangled mixed state (the so called  ``X-state") at finite $R$. The X-state reduces to a pure state (a singlet) in the $R\rightarrow 0$ limit while it saturates to an X-state with largest allowed value of von-Neumann entropy of $2 \ln2$ as $R\rightarrow \infty$ independent of the value of chemical potential. However, once an external magnetic field is applied to lift the spin-degeneracy, we find that the von-Neumann entropy of SSDM  becomes a function of the distance $R$ between the two spins. We also show that the von-Neumann entropy of TSDM in the $R\rightarrow \infty$ limit becomes a function of the chemical potential and it saturate to $2 \ln2$ only when the band in completely filled unlike the spin-degenerate case. Finally we extend our study to include spin-orbit coupling and show that it does effect these asymptotic results. Our findings are in sharp contrast with previous works which were based on continuum models owing to physics which stem from the lattice model.



\end{abstract}
\maketitle
\section{Introduction}{\label{sec:I}}

In recent years, significant effort has gone into developing understanding of quantum condensed matter system from a quantum information perspective~\cite{ Zeng2019,amico2008entanglement,gu2005ground,gu2004entanglement,larsson2005entanglement,vidal2006concurrence}. One of the key ingredients that has been used to  characterize quantum many-body systems is quantum ``entanglement" which has no classical counterpart \cite{horodecki2009quantum,o2001entangled, arnesen2001natural, wang2001entanglement, osborne2002entanglement, osterloh2002scaling, vidal2003entanglement, glaser2003entanglement} and a variety of idea involving 
von-Neumann entropy, concurrence~\cite{wootters1998entanglement}, mutual information~\cite{vedral2004mean,gu2007universal} etc. have been introduced to quantify the amount of entanglement in them. In particular quantum many-body systems comprising of indistinguishable particles have been studied extensively using such tools\cite{schliemann2001quantum,wiseman2003entanglement, ghirardi2004general,zanardi2002quantum,shi2003quantum,friis2013fermionic,benatti2014entanglement,debarba2017quantumness,iemini2014quantumness,majtey2016multipartite,gigena2017bipartite} and the phenomenon of quantum phase transition in such system has all been characterized using these ideas\cite{islam2015measuring, osterloh2002scaling, yuan2018thermodynamic,demokritov2006bose}. 

However, finding exact many-body states in case of fermions is a formidable task in general except for mean field theories where an exact Green's function approach can be used for systematic investigation. The framework of probing  non-interacting fermionic systems in terms of density matrices  was laid down by Dirac \cite{dirac1931note}, Lowdin \cite{lowdin1955quantum,lowdin1955quantum2}. Vederal~\cite{vedral2003entanglement}  studied  spin correlation of two electrons, located at different positions in terms of  entanglement in a system of non-interacting fermions, which Kim et.al  \cite{oh2004entanglement} then explored further. It was shown in \cite{oh2004entanglement} that for a continuum model of free fermions in three-dimensions, the entropy of two spin degrees of freedom (TSDM) varies with the distance $R$ between them and it saturates to its maximum value of $2 \ln2$ as $R\rightarrow \infty$. However, the single spin density matrix (SSDM) obtained by partial tracing the TSDM is always found to be maximally mixed (von-Neumann entropy being $\ln 2$) and hence is independent of the distance $R$ between the two spins.  We show that, once the fermions are placed on a lattice and an external magnetic field is applied to break the spin degeneracy, the entropy of SSDM starts to depend upon the distance $R$ between the two spins. Moreover, the saturation value of entropy of TSDM in the $R\rightarrow \infty$ limit is no longer $2 \ln2$ rather it depends upon the chemical potential of the system. These features are indicative of physics beyond what is studied in  \onlinecite{oh2004entanglement} i.e., effects that can not be captured by a spin degenerate Fermi sea for a continuum model. Motivated from this fact we study a spin correlation encoded in TSDM and SSDM using this framework given in Ref.[\onlinecite{oh2004entanglement}] for a one dimensional spinfull Fermionic chain which is subjected both Zeeman field and spin-orbit field, as a function of the chemical potential.\\  
\indent The remainder of this article is organized as follows. In Sec.~\ref{secII}, we introduce our 1D lattice model and define TSDM including both Zeeman and Rashba spin-orbit coupling (RSOC) interaction terms and  analyse the TSDM entropy for the case where both terms are zero.  In Sec.~\ref{secIII}, we study the case where only the  Zeeman coupling ($B\neq0$ and $\lambda=0$) is present. We examine the variation of entropy of TSDM as well as SSDM as a function of the chemical potential. In Sec.~\ref{secIV} we first study entanglement in the presence of the RSOC term alone followed by the case where both terms are present ($B, \lambda \neq0$). Finally, we conclude in  Sec.~\ref{secVI}.


\section{Model and method \label{secII}}
We begin with the 1D tight-binding model Hamiltonian in real space, which takes the following form :

\begin{equation}
 \begin{split}
H &= -t \sum_{i}^{M} ( c_{i}^{\dagger}  c_{i+1} + h.c. )  + B \sum_{i}^{M}  c_{i}^{\dagger}   \sigma_{x} \   c_{i} \\
                         &~~~~~~~~~~~   + i \ \lambda \sum_{i}^{M} ( c_{i}^{\dagger} \sigma_{y} c_{i+1}+ h.c)\ ,
\end{split} 
\label{E1}
\end{equation}

where $t$ is the hopping parameter, $B$ and $\lambda$ are the Zeeman and RSOC strengths respectively. Within our analysis, the value of $B$ and $\lambda$ have been chosen in terms of the hopping parameter $t$ which is fixed to $t=1$. Also, the distance scale $R$ is set in terms of the lattice spacing $a$, throughout our analysis. 

We consider periodic boundary condition (PBC) and work in momentum space to obtain the single-particle spectrum. The corresponding dispersion relation $E_{\pm}(k) = -2 t \cos(k a)\pm \sqrt{B^{2} + 4 \lambda^{2} \sin^{2}(k a)} $ defines two bands with the following eigen-functions,

\begin{equation}
\phi_{-,k}(r) =\frac{e^{- i k r}}{\sqrt{2 L}}  \begin{pmatrix}
 - e^{-i \theta_{k}}\\ 
1
\end{pmatrix} ; \   \phi_{+,k}(r) = \frac{e^{- i k r}}{\sqrt{2 L}}  \begin{pmatrix}
  e^{-i \theta_{k}}\\ 
1
\end{pmatrix}\ ,
\label{E2}
\end{equation}{}
with $L = Ma$.  Here $a$ is the lattice constant. Also, $e^{- i \theta_{k}} = |Z|/Z $, where $Z$ is defined as $Z = B + 2i \lambda \sin(ka)$. 

In an earlier work, L\"{o}wdin proposed the idea of fundamental invariant~\cite{lowdin1955quantum2} employing which, density matrix of any order for a given many-body wave-function can be obtained 
and written as follows:

\begin{equation}
\rho^{}(x_1,x_2)= \sum_{k l} \phi^{*}_{\pm,k}(r_{1}, \sigma_{1})  \ \phi_{\pm,l}(r_{2},\sigma_{2})\ ,
\label{E3}
\end{equation}{}

\noindent

where, $x= (r, \sigma)$ denotes position and spin quantum number of electron, and $\sigma= (\uparrow / \downarrow)$ represents up and down components of $\phi_{\pm,k}(r)$. Using the fundamental invariant, one can write elements of two particle density matrix as~\cite{lowdin1955quantum,lowdin1955quantum2}
\begin{equation}
\rho^{(2)}(x_1,x'_1,x'_2,x_2)=\frac{1}{2}
\begin{vmatrix}
 \rho^{(1)}(x'_{1},x_{1})  & \rho^{(1)}(x'_{1},x_{2})  \\ 
  \rho^{(1)}(x'_{2},x_{1}) & \rho^{(1)}(x'_{2},x_{2})
\end{vmatrix}\ .
\label{E4}
\end{equation}{}

Here, superscript $(1)$ and $(2)$ denote one and two particle density matrix elements respectively. Since our model consists of two bands and therefore have two different Fermi momenta $k_{f}^{-}$ 
(for lower band) and $k_{f}^{+}$ (for upper band), for any arbitrary chemical potential denoted as $\delta$. The single particle density matrix elements in Eq.(\ref{E4}) namely, $\rho^{(1)}(x,x')$ can then be 
defined as:

\begin{equation}
 \begin{split}
\rho^{(1)}(x,x') &=\sum_{|k|=k_{i}}^{k_{f}^{-}} \phi^{*}_{-,k}(r,\sigma)  \ \phi_{-,k}(r',\sigma') \\
                         &~~~~~~~~~~~    + \sum_{|k|=0}^{k_{f}^{+}} \phi^{*}_{+,k}(r,\sigma)  \ \phi_{+,k}(r',\sigma')\ ,
\end{split} 
\label{E5}
\end{equation} 

where, Fermi momenta $k_{f}^{-}$ and $k_{i}$ can be obtained from the relation $\mu = E_{-}(k)$ and  momentum $k_{f}^{+}$ from the relation $\mu = E_{+}(k)$.  Temperature is assumed to be zero in our entire analysis. Therefore, the mean occupation number at each momentum below the Fermi level is unity. Since our objective in this article is to study the entanglement of the spin degrees of freedom, we obtain TSDM using 
Eq.(\ref{E4}). We also consider diagonal elements of space density matrix: $r_{1}= r_{1}'$ and $r_{2}= r_{2}'$ to compare with the earlier results~\cite{vedral2003entanglement, oh2004entanglement}  
and have physical interpretation of elements with $\sigma=\sigma'$ as probabilities~\cite{lowdin1955quantum2,dirac1931note}. Hence we define generic element of TSDM $\rho^{(2)}_{\sigma_{1},\sigma_{2};\sigma_{1}',\sigma_{2}'}$ using Eq.(\ref{E4}) as

\begin{equation}
 \begin{split}
\rho^{(2)}_{\sigma_{1},\sigma_{2};\sigma_{1}',\sigma_{2}'} &=\frac{1}{2}\ [ \  \rho^{(1)}(r_{1} \sigma_{1},r_{1} \sigma_{1}')  \  \rho^{(1)}(r_{2} \sigma_{2},r_{2} \sigma_{2}')
 \\
                         &~~~~~~~~  -   \rho^{(1)}(r_{1} \sigma_{1},r_{2} \sigma_{2}') \   \rho^{(1)}(r_{2} \sigma_{2},r_{1} \sigma_{1}') \ ]\ ,
\end{split} 
\label{E6}
\end{equation} 
where, each single particle density matrix can be written in terms of spin-orbital wave function $\phi_{\pm,k}(r)$ using  Eq.(\ref{E5}). We thus obtain the full TSDM, which can be written as 

\begin{widetext}
\begin{equation}
\rho^{(2)}_{12}= N
\begin{bmatrix}
  m^{2}- G^{2}_{r}&  -m A + G_{r} H_{r} & -m A + G_{r} K_{r} &  A^{2} - H_{r} K_{r}\\ 
-m A+  G_{r} H^{*}_{r}  &  m^{2}- H_{r}H^{*}_{r} & A^{2}- G^{2}_{r}  &  -m A + G_{r} H_{r}\\ 
-m A + G_{r} K^{*}_{r} &  A^{2}- G^{2}_{r}  &   m^{2}- K_{r}K^{*}_{r} &  -m A + G_{r} K_{r} \\ 
A^{2} - H^{*}_{r} K^{*}_{r} &   -m A + G_{r} H_{r} &  -m A + G_{r} K^{*}_{r}  & m^{2}- G^{2}_{r}
\end{bmatrix}\ .
\label{E7}
\end{equation}
\end{widetext}

In Eq.(\ref{E7}), $N$ is the normalization constant and is equal to inverse of sum of the diagonal elements $i.e.$ $1/N= 4 m^{2}- 2 G^{2}_{r} - H_{r}H^{*}_{r} - K_{r}K^{*}_{r}$. Also, $m = \sum_{|k|=k_{i}}^{k_{f}^{-}} 1+  \sum_{|k|=0}^{k_{f}^{+}} 1$; $A= \sum_{|k|=k_{i}}^{k_{f}^{-}}  e^{i \theta_{k}}- \sum_{|k|=0}^{k_{f}^{+}}  e^{i \theta_{k}}$
and the functions  $G_{r} =\sum_{|k|=k_{i},0}^{k_{f}^{-},k_{f}^{+}}  e^{{i  k (r_{1}-r_{2})}}$, $H_{r} =\sum_{|k|=k_{i}}^{k_{f}^{-}} \  e^{i \theta_{k}} \  e^{{i  k (r_{1}-r_{2})}} - \sum_{|k|=0}^{k_{f}^{+}} \  
e^{i \theta_{k}} \  e^{{i  k (r_{1}-r_{2})}}$ and $K_{r} =\sum_{|k|=k_{i}}^{k_{f}^{-}} \  e^{i \theta_{k}} \  e^{{i  k (r_{2}-r_{1})}} - \sum_{|k|=0}^{k_{f}^{+}} \  e^{i \theta_{k}} \  e^{{i  k (r_{2}-r_{1})}}$. Here, it should be noted that the  functions $A$, $K_{r}$ and $H_{r}$ are such that they are zero when there is no spin-distinguishing terms $i.e.$  $B=\lambda=0$. Henceforth, we use the term "spin-pair" to denote the pair of spins in the TSDM. We now define the filling of the system by imposing the following constraint equation in the mean field level given by:

\begin{equation}
<n_{i,\uparrow}>+<n_{i,\downarrow}>=\delta  ,
\label{Ec}
\end{equation}

where $\delta$ carries the information about average filling at each lattice site $i$ and fixes the chemical potential. 

We first consider the case when both $B=0$ and $\lambda=0$. This corresponds 
to $k_{f}^{-}=k_{f}^{+}$ and $k_{i}=0$. This in turn leads to double degeneracy of the eigenspectrum. As a result, the spin-distinguishing functions in the expression of TSDM (Eq.(\ref{E7})) reduces to $A=H_{r}=K_{r}=0$. Hence, the TSDM in Eq.(\ref{E7}) reads as:

\begin{equation}
\rho^{(2)}_{12}= \frac{1}{2(2m^{2} - G_{r}^{2})}
\begin{bmatrix}
  m^{2}- G^{2}_{r}&   0 & 0 &   0\\ 
0  &  m^{2} & - G^{2}_{r}  & 0\\ 
0 &  - G^{2}_{r}  &   m^{2} &  0 \\ 
0 &   0 &  0  & m^{2}- G^{2}_{r}
\end{bmatrix}\ .
\label{E8}
\end{equation}{}

\begin{figure}[htb!]
\centering
\vspace{0.2cm}
\includegraphics[width=6.0cm, height=5.0cm]{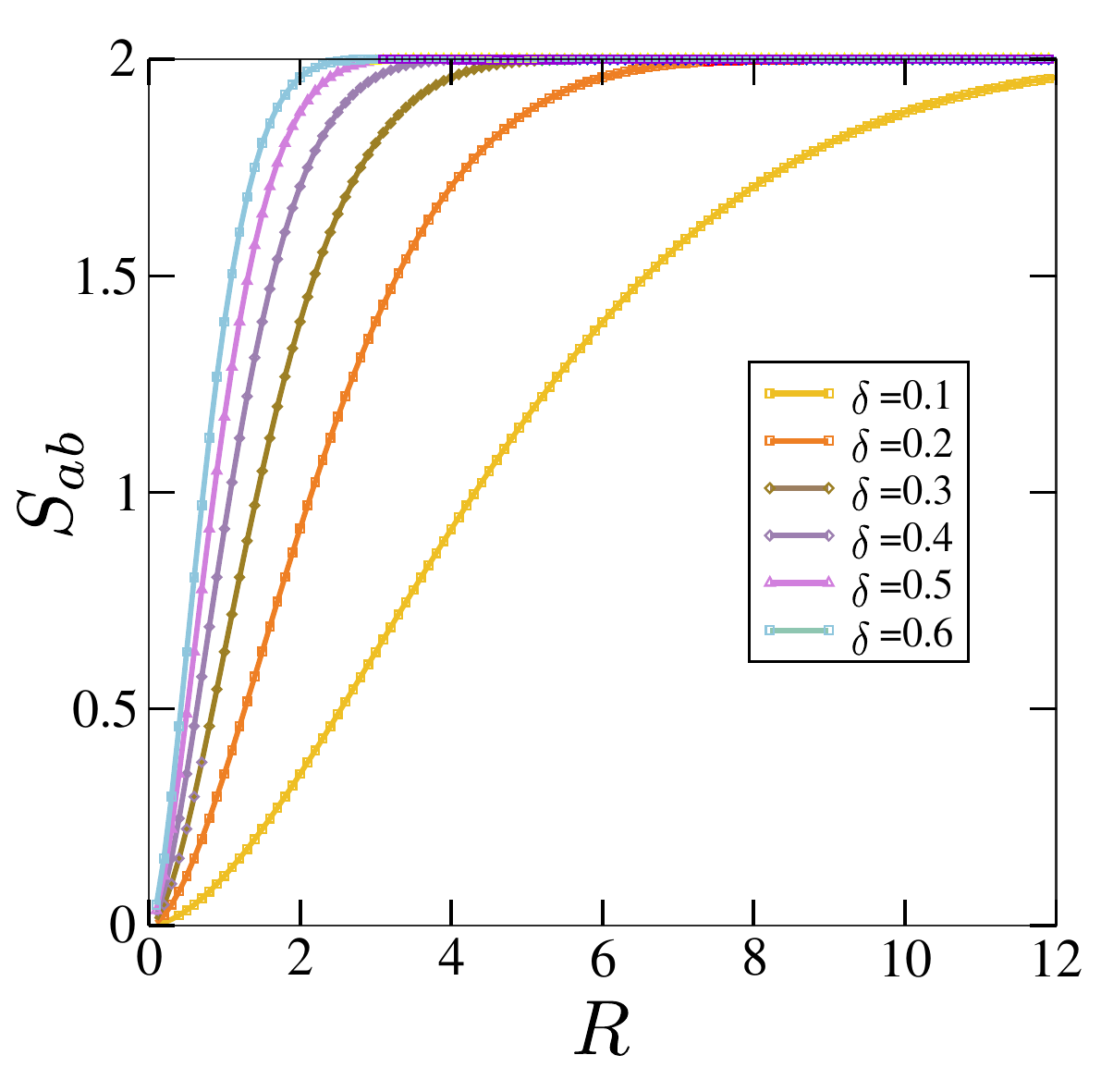} \vspace{-0.2cm}
	\caption{(Color online) 
	The variation of entropy (in the units of $\ln 2$)  for the   case $B=\lambda=0$ is depicted as a function of distance $R= |r_{1}- r_{2}|$ for different filling $\delta$, starting from $\delta$=0.1 (yellow) to $\delta=0.6$ (cyan). The number of sites has been taken to be $M= 500$ for this plot as well as all the plots that follow.}
\label{f0}
\end{figure}

The TSDM, given by Eq.(\ref{E8}), only represents a pair of spin-1/2 degrees of freedom  residing at a distance $R$ from each other and the entropy corresponding to this density matrix indicates the entanglement of two spins with the rest of the Fermi sea. The two spin state in the Eq. (\ref{E8}) are known as maximally entangled mixed states or popularly known as ``X- states" \cite{frank2001}. In Fig.~\ref{f0}, we illustrate the behavior of entropy $S_{ab}$, corresponding to the TSDM $\rho_{12}^{(2)}$, as a function of distance $R$ between them for various fillings $\delta$ ranging from (0.1 to 0.6). For $R\rightarrow 0$ entropy $S_{ab}\rightarrow 0$ implying  a complete decoupling of the spin-pair from the rest of the Fermi sea. In fact, the TSDM reduces to a spin-singlet. However, in the large $R$ limit entropy  $S_{ab}$ approaches the asymptotic value of $2 \ln 2$ independent of $\delta$ though  the rate  ($d S_{ab}/d R $ ) at which it reaches asymptotic value is greater for larger value of $\delta$. Note that  $2 \ln 2 $ is the maximum value of allowed entropy for a pair of spin-1/2 and  corresponding states having this maximum value are known as ``2-entangled states" \cite{scott,parisi,cerf, sowrabh}.

\section{Zeeman field and entropy reduction \label{secIII}}
In this section we analyse the case where only  $B \neq 0$.  This lifts the spin degeneracy and breaks the time reversal symmetry. We find that when  $B \neq 0$ and  RSOC term  $\lambda=0$, then the factor 
$e^{i \theta_{k}}=1$, which results in following constraints on the functions in TSDM (Eq.(\ref{E7})): $ H_{r} = K_{r} = H^{*}_{r} = K^{*}_{r}$. The un-normalized TSDM $\rho^{(2)}_{12}$ then reduces to:

\begin{widetext}
\begin{equation}
\begin{bmatrix}
  m^{2}- G^{2}_{r} & -m A +G_{r} H_{r} & -m A+ G_{r} H_{r}& A^{2}-  H_{r}^{2}\\ 
 -mA+G_{r} H_{r}  &  m^{2}- H_{r}^{2} & A^{2} - G^{2}_{r}  & -m A+ G_{r} H_{r}\\ 
-m A+G_{r} H_{r} &  A^{2}- G^{2}_{r}  &   m^{2}- H_{r}^{2} & -mA+  G_{r} H_{r} \\ 
  A^{2}-H_{r}^{2}  &  -m A + G_{r} H_{r} & -mA + G_{r} H_{r}  & m^{2}- G^{2}_{r}
\end{bmatrix}\ ,
\label{E9}
\end{equation}{}
\end{widetext}

where we have left the factor $1/(2 L)^{2}$. We now compute the entropy corresponding to this TSDM for different chemical potentials.  We begin by filling the lower energy band represented by the dispersion relation $  E_{-}(k) = -2 t \cos(k a) - \sqrt{B^{2} + 4 \lambda^{2} \sin^{2}(k a)} $.  In this case the functions in TSDM can be further simplified as: $G_{r} = H_{r}$ and $A=m$. Therefore, the TSDM is a pure state given by the  triplet $\ket{\psi_{t_{1}}} = \ket{00}$ (where $\ket{0}$ is one of the eigenstate of Pauli-x matrix) and is independent of $R$ as expected.
\begin{figure}[htb]
\centering
\includegraphics[width=8.75cm, height=7.5cm]{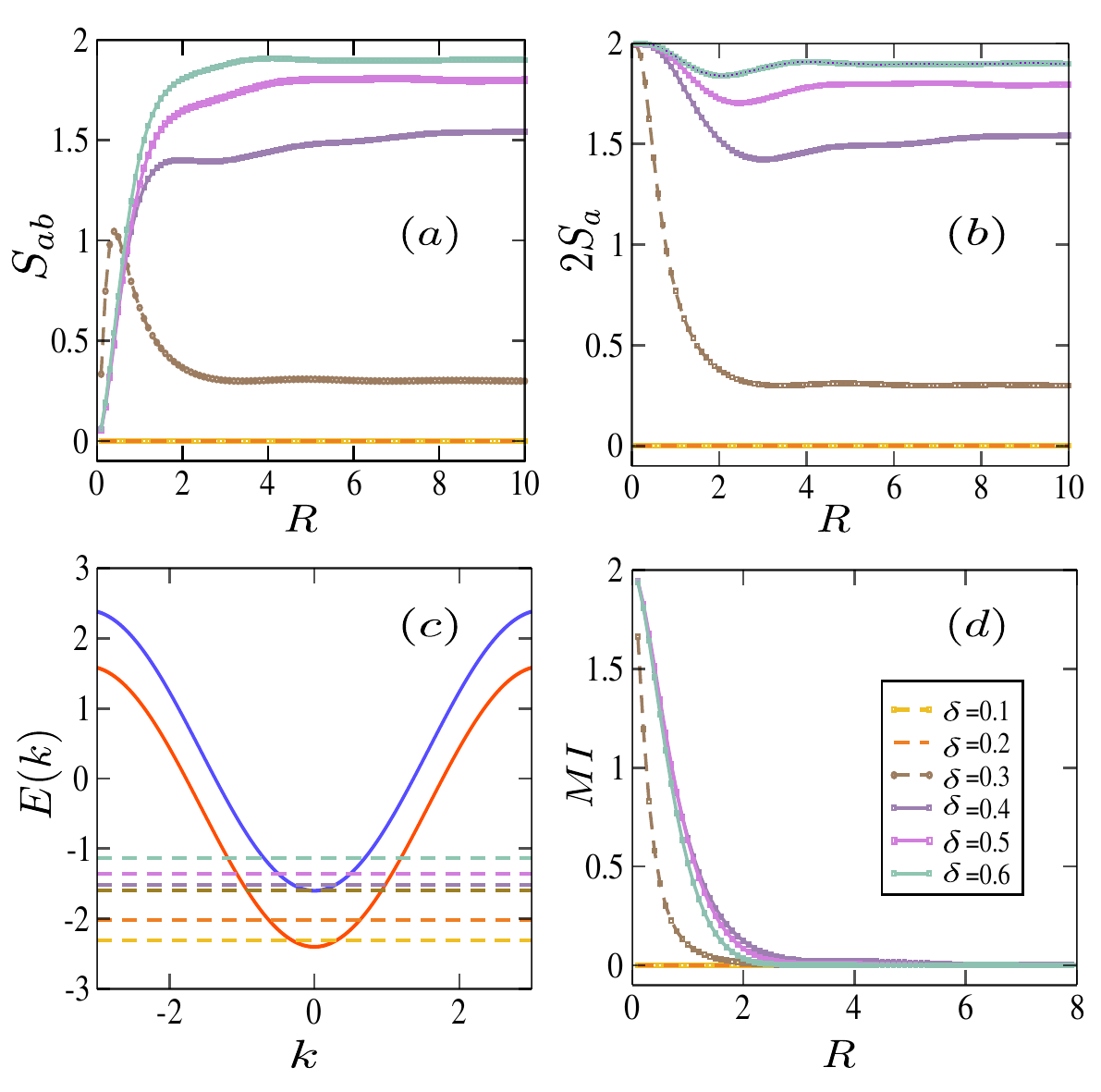}
	\caption{(Color online) (a) This diagram demonstrates the variation of entropy (denoted as $ S_{ab}/\ln 2$) of the TSDM $\rho_{12}^{(2)}$ (Eq.(\ref{E9})) as a function of distance $R= |r_{1}- r_{2}|$ 
	for different filling $\delta$, starting from $\delta$=0.1 (yellow) to $\delta=0.6$ (cyan).  Here we choose $B=0.4$ and RSOC term $\lambda=0$. $B$ is chosen in terms of the hopping parameter $t$. (b) Variation of entropy of SSDM is shown. (c) Corresponding fillings in the band structure are included. (d) MI for different fillings $\delta$ as a function of the distance between the two spins.}
\label{f1}
\end{figure}

\begin{figure}[htb]
\centering
\includegraphics[width=8.75cm, height=7.5cm]{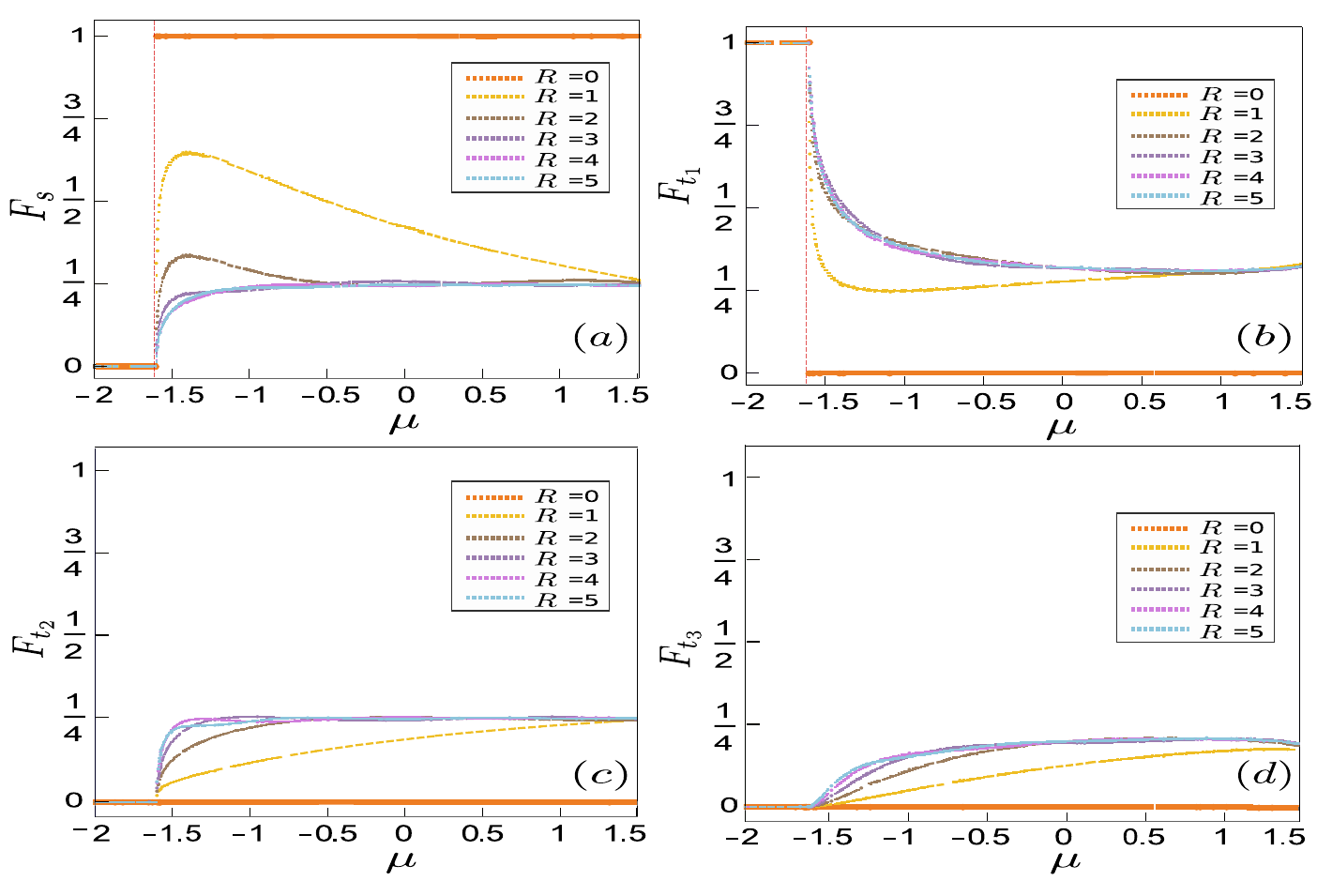}
	\caption{(Color online) Fidelities of TSDM in Eq. (\ref{E9}) with the spin- singlet $\ket{\psi_{s}}$ and three triplets  $\ket{\psi_{t_{1}}}$, $\ket{\psi_{t_{2}}}$ and $\ket{\psi_{t_{3}}}$ are illustrated respectively in (a), (b), (c) and (d) for varying chemical potential $\mu$. The vertical red dashed line represents the value of chemical potential $\mu$ ($\delta \approx 0.3$) for which the fermions start to fill  the upper band. Value of magnetic field is $B=0.4$ and of RSOC parameter is  $\lambda=0$.}
\label{f1p}
\end{figure}



However, when we fill the upper band $E_{+}(k)$ with electrons having opposite spin polarisation w.r.t. the lower band then the above-mentioned relations namely, $G_{r} = H_{r}$ and $A=m$ do not hold, leading to a TSDM which is a mixed state (Fig. \ref{f1} (a)). This implies that now the spin-pair is entangled with the rest of the Fermi sea. We analyse this entanglement as a function of distance  $R$ and chemical potential $\mu$, where $\mu$ is tuned by varying the filling fraction $\delta$. It is interesting to note that even when we add a single electron in the upper band ($\delta \approx 0.3$) while the lower band is partially filled, TSDM undergoes a triplet $\ket{\psi_{t_{1}}} = \ket{00}$ to singlet $\ket{\psi_{s}} = (1/\sqrt{2}) \ (  \ket{10} - \ket{01} )$ transition in case $R =0$ as clear from the Fig. \ref{f1p} (a)-(d), where the four fidelities $F_{s}$ and $F_{t_{i}}$s are defined as : $F_{s} = \bra{\psi_{s}} \rho_{12}^{(2)} \ket{\psi_{s}}$ and $F_{t_{i}} = \bra{\psi_{t_{i}}} \rho_{12}^{(2)} \ket{\psi_{t_{i}}}$. Here, the density matrix $\rho_{12}^{(2)}$ is the TSDM (normalized) in Eq. (\ref{E9}) and $\ket{\psi_{t_{1}}} = \ket{00}$, $ \ket{\psi_{t_{2}}} =(1/\sqrt{2}) \ (  \ket{10} + \ket{01} )$ and $\ket{\psi_{t_{3}}} = \ket{11}$. This transition exemplifies remarkable sensitivity of the entropy to the degeneracy of fermionic states with opposite spins. Moreover, in sharp contrast with  the study of spin degenerate free fermionic continuum model in \cite{oh2004entanglement} and the spin-degenerate case on the lattice model here ($B, \lambda = 0$), the saturating value of entropy of TSDM  in large $R$ limit has strong dependence on the filling fraction $\delta$. This saturating value can never reach the maximum entropy value of $ 2 \ln 2$, unless all the states in the two bands are filled $i.e.$ $\delta=2$ (Fig. \ref{f1} (a)). In this case all four states namely singlet and triplets have equal occupancy $i.e$ four fidelities $F_{s}$ and $F_{t_{i}}$s are equal to $1/4$ independent of the value of the distance $R$, other than when $R=0$ ($i.e.$ two spins at the same site).


Another feature which is in contrast with the results in \cite{oh2004entanglement} is the tunability of  the entropy of a single spin with the rest of the Fermi sea by the chemical potential and distance $R$, in the presence of magnetic field. In order to examine this, we again consider  a situation where the lower band is partially filled and the upper band is empty. In this case the normalized reduced SSDM $\rho_{1} $ turns out to be a pure state and reads as $\rho_{1} = (1/2) \ (\mathbb{I} - \sigma_{x})$, where $\sigma_{x}$ is a Pauli-$x$ matrix. To explore the single spin entropy behaviour as the upper band $E_{+}(k)$ is filled, we first note that the constraints on the functions in the TSDM change as: 
$ H_{r} = K_{r} = H^{*}_{r} = K^{*}_{r}$, though the factor $e^{i \theta_{k}}=1$ remains unchanged. These constraints in turn, simplifies the normalized SSDM $\rho_{1}$, which acquires the following form:

\begin{equation}
\frac{1/(2 L)^{2}}{\rm{Tr}(\rho^{(2)}_{12})}
\begin{bmatrix}
 2m^{2}- G^{2}_{r}- H_{r}^{2} &  -2 m A + 2 G_{r}H_{r}\\ 
 -2 m A + 2 G_{r}H_{r} &  2m^{2}- G^{2}_{r}- H^{2}_{r}
\end{bmatrix}\ .
\label{E10}
\end{equation}{}


It is evident that the SSDM $\rho_{1}$ (Eq.(\ref{E10})) above, depends upon the distance between the two spins via TSDM $\rho^{(2)}_{12}$.  More interestingly, 
$\rho_{1}$ is not a pure state anymore rather a maximally mixed state when the distance between the spins is zero \ie $R =|r_{1}-r_{2}|=0$. The off-diagonal elements in the expression of $\rho_{1}$ in this case, 
turns out to be zero as $G_{r} = m$ and $H_{r}= A$, relations still hold when two spins are located at the same site within the chain. Also, since the $\rm{Tr}(\rho^{(2)})$ = $2 (m^{2}- A^{2})/(2L)^{2}$, 
each of the diagonal elements becomes $1/2$.  We now focus on the profile of entropy in the large $R$ limit and is depicted in Fig.~\ref{f1}(b). Note that, when the chemical potential $\mu$ lies in the lower band $E_{-}(k)$, the entropy corresponding to SSDM is zero and independent of the distance $R$ (see Fig.~\ref{f1}(b) with filling values $\delta=0.1$ and $\delta=0.2$). However, when we start to fill the upper band $E_{+}(k)$ as well, we observe that entropy of SSDM saturates to non-zero value which increases upon increasing the filling values from $\delta=0.3$ 
to $\delta=0.6$, and finally saturates to the maximum value of $ \ln 2$ when both the bands are completely filled.  The corresponding fillings in the band structure are indicated in Fig. (\ref{f1} (c)).

\begin{figure}[htb!]
\centering
\includegraphics[width=9.0cm, height=7.5cm]{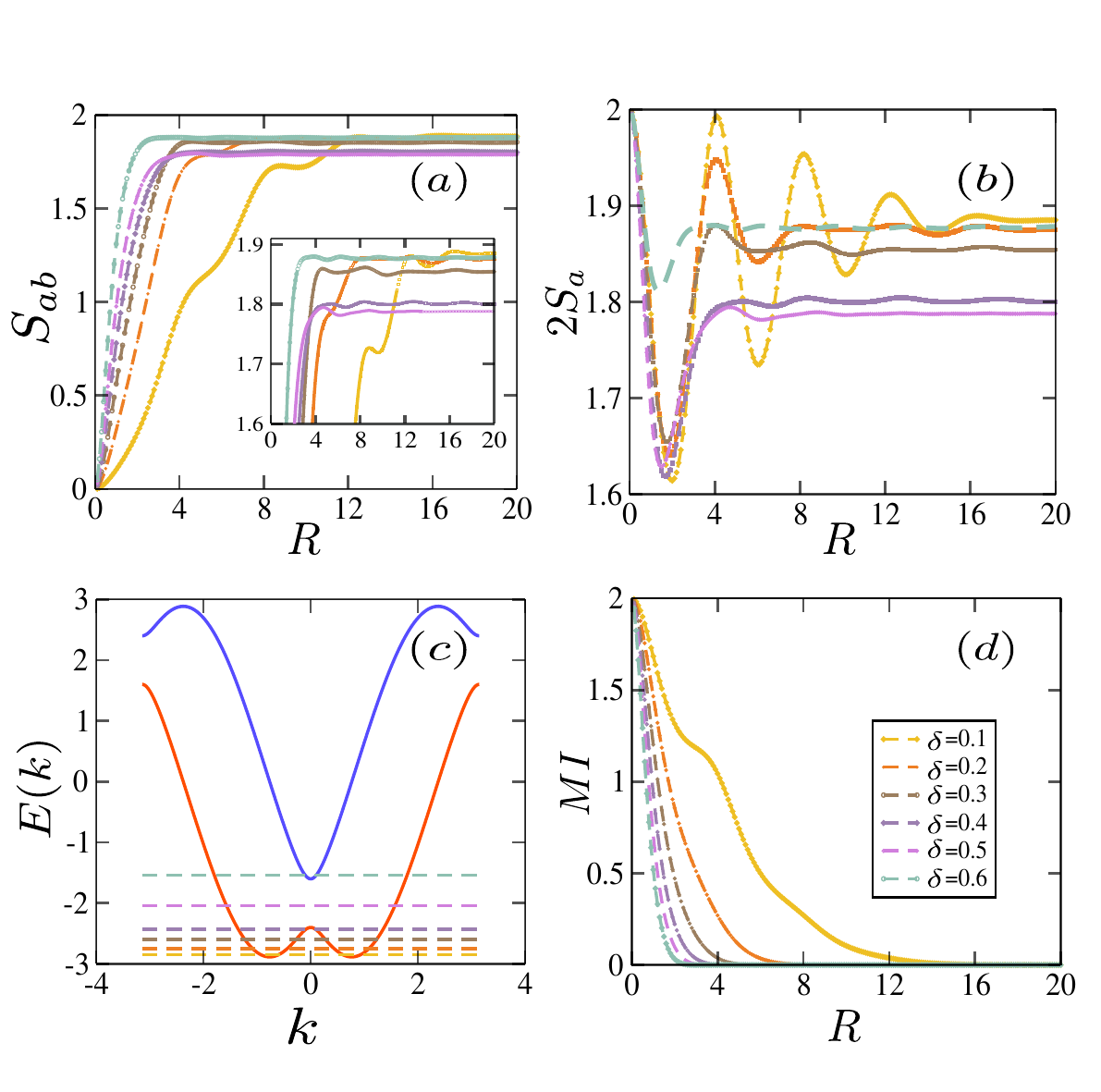}
	\caption{(Color online) In panel (a) we depict the variation of entropy of the TSDM as a function of distance for various filling starting from yellow ($\delta$=0.1) to the cyan ($\delta$=0.6). 
	In panel (b) we show the variation of SSDM as a function of distance $R$ for fillings ($\delta$=0.1 - 0.6) respectively. (c) The spectrum of our 1D chain along with different fillings $\delta$ is shown in panel (c). The spatial dependence of MI is sketched in the panel (d). See text for discussion.  We choose $\lambda$=1.0, $B=0.4$ for all the panels.
	}
\label{f5}
\end{figure}


Now we investigate the  degree of correlation between the two spins in TSDM by studying mutual information (\rm {MI}), which is defined as $\rm {MI} = 2 S(\rho_{1})-S(\rho^{(2)}_{12})$. Here $S(X)$ denotes the entropy of a given density matrix $X$. Note that MI is zero for the product state of two spins (hence uncorrelated), but it can also be zero while $S(\rho_{1})= 1/2  \ S(\rho^{(2)}_{12})$ $i.e.$ when the total entropy follows an addition law of individual entropies and in this case also spins are uncorrelated. In Fig.~\ref{f1}(d) we demonstrate the MI variation  as a function of distance  $R$ for various fillings ranging from $\delta=0.1$ 
to $\delta=0.6$. We find that MI is a monotonically decreasing function of distance $R$ for the fillings $\delta =0.3-0.6$ and identically zero for $\delta =0.1-0.2$. Note that, the corresponding behavior of the entropies is non-monotonic with respect to the filling factor $\delta$. As before, when we fill only the lower energy band $E_{-}(k)$, MI is zero as TSDM is a pure state which is also a product state in this case. However, once we put a single electron in the upper band $E_{+}(k)$, MI  attains the maximum value of $2 \ln 2$ at $R = 0$. Interestingly, as the distance between the two spins $R$ is increased, MI decays monotonically to zero again in large $R$ limit. Although, both the single spin (represented by SSDM) and two spins (represented by TSDM), that have been taken into consideration are highly entangled with the rest of the Fermi sea and are maximally entangled in case $\delta=2$, there is no correlation between the two spins in the spin-pair for large distance $R$.

\section{Rashba SOC and Zeeman field \label{secIV}}
In this section, we first turn off the Zeeman term \ie $B=0$, and assumes only RSOC $\lambda \neq 0$. This implies that there is a double degeneracy owing to the time-reversal symmetry at $k=0$. The factor $e^{\pm i \theta_{k}} = \pm i$ for $B = 0$ implies $e^{ i \theta_{-k}} = e^{- i \theta_{k}}$. Hence, the constraints on the functions within the TSDM are : $A=0 ; H_{r}= -K_{r} = H^{*}_{r} = -  K^{*}_{r}=0$. Moreover, the functions $m$ and $G_{r}$ remain invariant, leading to TSDM which has exactly the same functional form as given in Eq. (\ref{E8}) $i.e.$ retains the same functional form of ``X-state", which is not the case for $B \neq 0$ and $\lambda=0$ as discussed in Sec.\ref{secIII}. Moreover, as discussed the entropy of a single spin is again reduce to its maximum value of $\ln 2$ independent of $R$ and chemical potential. Therefore, it is apparent that even in presence of non-zero $\lambda$, the entropy of TSDM as well as SSDM remains unchanged $i.e.$ are matching with the case where both $B =0, \lambda=0$.  However, once we switch on the Zeeman term $B \neq 0$ as well,
the resulting  TSDM do not confine to the form of ``X-states" and is given by Eq. (\ref{E7}). The entropy corresponding to TSDM is depicted in Fig.~\ref{f5}(a). In the limiting case $R=0$, TSDM is yet a pure state independent of filling fraction $\delta$. In order to find which pure state TSDM is in when $R=0$ we write the entropy of a single spin in presence of both $B$ and $\lambda$. The SSDM $\rho_{1}$ (un-normalized) for this case can be written as :

\begin{equation}
\begin{bmatrix}
 2m^{2}- G^{2}_{r}- H_{r}^{2} &  -2 m A + G_{r}(K_{r} + H_{r})\\ 
 -2 m A + G_{r}(K_{r} + H_{r}) &  2m^{2}- G^{2}_{r}- K_{r}^{2}
\end{bmatrix}\ , 
\label{E12}
\end{equation}{} 

the entropy reaches maximum value of $\ln 2$  when distance $R=|r_{1}-r_{2}|=0$. This confirms the fact that TSDM  not a product state, rather a maximally entangled state. It should be noted that when the chemical potential is such that only the lower band is filled $i.e.$ $\delta$ ranging from $0.1-0.5$, still the TSDM is not a triplet $\ket{\psi_{t_{1}}}$, unlike the case where only $B \neq 0$ and $\lambda=0$ in Sec. \ref{secIII}. In fact, we find that TSDM turns out to be a singlet state $\ket{\psi_{s}}$. Another interesting observation to note is that the entropy of SSDM oscillates with a decaying envelope, as a function of distance $R$, owing to the competition of Zeeman term $B$ and the RSOC term $\lambda$.  Finally, in Fig.~\ref{f5}(d) we demonstrate the behavior of $\rm MI$ as a function of distance $R$ for various fillings ranging from $\delta=0.1$ to $\delta=0.6$. 
We observe that for filling $\delta$=0.1-0.5, $i.e.$ when the chemical potential lies in the lower band, MI is maximum at $R=0$ and gradually suppressed over distance in a monotonic decaying fashion unlike the case discussed in Sec. \ref{secIII} where MI is identically zero independent of $R$ when only the lower band is filled. Moreover, as we increase the filling gradually,  $\rm MI$ exhibits faster decay rate over distance, which implies that the correlations between the individual spin  in the spin-pair and the rest of Fermi sea  grow stronger as the depth of the Fermi sea is increasing, and hence the faster decay.

\begin{figure}[htb!]
\centering
\includegraphics[width=8.9cm, height=4.25cm]{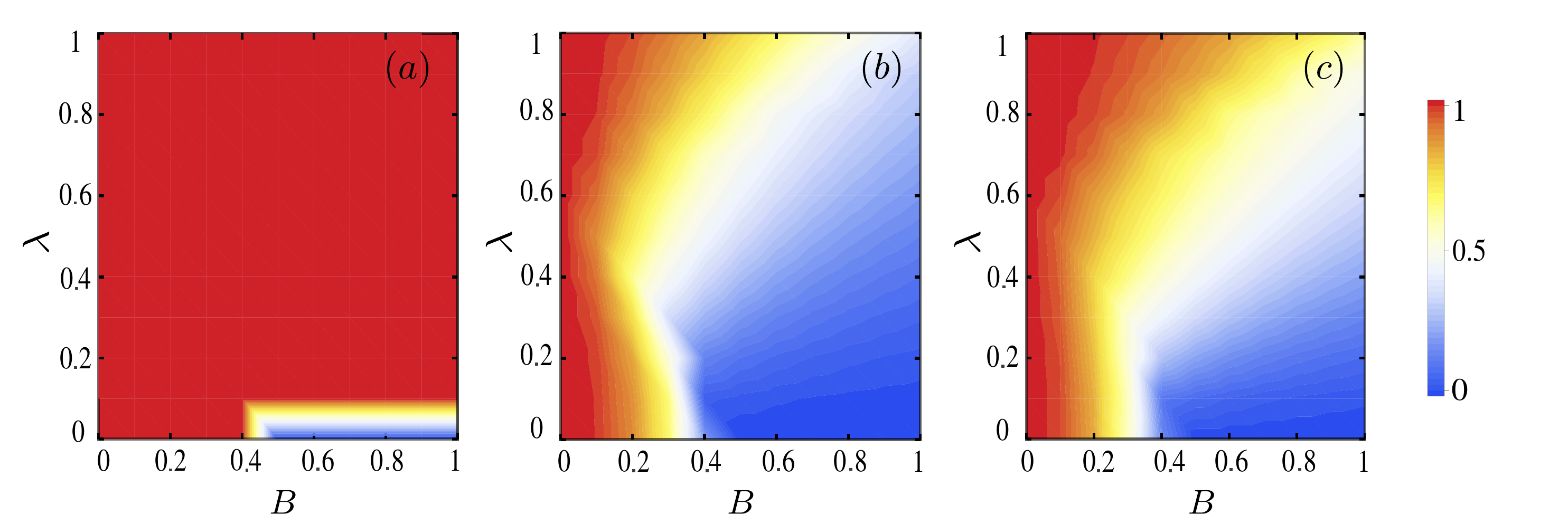}
\caption{(Color online) Entropy of SSDM   $ S_{a}/\ln 2$ demonstrated in the plane of magnetic field $B$ and RSOC $\lambda$ in the three panels  (a), (b) and (c), where the distance $R$  has been set to 0, 2 and 10 respectively.  Here blue color corresponds to a SSDM which is a pure state whereas  red denotes a SSDM which is a maximally mixed state ($\rm ln~2$) state. We choose the filling fraction $\delta$ to be $\delta$=0.3 in all the three panels. }
\label{f6}
\end{figure}

\section{Conclusion\label{secVI}}

To conclude, we employ spin density matrix approach proposed by Kim etal. (\cite{oh2004entanglement}) to analyse the many-body fermionic states of a 1-D lattice model in the presence of magnetic field and RSOC. In contrast to their study, we find that SSDM is a function of the distance between the two spins in the spin-pair once the magnetic field is turned on. Moreover, TSDM can not get maximally entangled $i.e$ entropy is $ 2 \ln 2$ with the rest of the Fermi sea in the large $R$ limit ($R \rightarrow \infty$), unless both the bands are completely filled. Note that this is also in sharp contrast with   the case in \cite{oh2004entanglement}, where the energy spectrum is unbounded above and has no well defined sense of filling fraction and hence the related physics does not exist in their continuum model. Finally, for the case where only magnetic field is present, we also note the fact that even when we add a single electron in the upper band while the lower band is partially filled, SSDM entropy  undergoes a sharp transition (corresponds to triplet to singlet transition of TSDM ) in the limit $R =0$  (see Fig. \ref{f6} (a)). This entropy transition persists for small distance corresponding to few lattice sites and for large $R$ as well, though the sharpness is reduced (see Fig. \ref{f6} (b),(c)). It is also evident from  Fig. \ref{f6} that  when $B=0$, SSDM  entropy $S_{a}$ is independent of the strength of RSOC like it was in the case studied in \cite{oh2004entanglement}.

\subsection*{Acknowledgements} 
A.V.V. acknowledges the Council of Scientific and Industrial Research (CSIR), Govt. of India for financial support. S.D. would like to acknowledge the MATRICS grant (Grant No. MTR/ 2019/001 043) from the Science and Engineering Research Board (SERB) for funding.

\vspace{+0.2cm}
\emph{Author contribution:-}
The first two authors, S.J.  and A.V.V. have contributed equally to this work. \\

\vspace{+0.2cm}

\bibliography{bibfile}{}

\vspace{+0.9cm}

\end{document}